\documentclass[aps,pra,twocolumn,preprintnumbers,amsmath,amssymb,10pt]{revtex4-1}
\usepackage{times}

\usepackage[usenames,dvipsnames]{xcolor}

\newcommand{\GGrev}[1]{\textcolor{black}{#1}}

\usepackage[english]{babel}
\usepackage{graphicx}
\usepackage{dcolumn}
\usepackage{bm}
\usepackage{color}
\usepackage{amsmath}
\usepackage{relsize,amsmath,dsfont,mathrsfs,empheq,verbatim,upgreek}
\usepackage{etoolbox}
\usepackage[caption = false]{subfig}
\apptocmd{\thebibliography}{\raggedright}{}{}

\usepackage[colorlinks=true,linkcolor=blue,urlcolor=blue,citecolor=blue,pdfusetitle]{hyperref}
\usepackage{graphics}
\usepackage{bm}
\usepackage{color}
\usepackage{amscd}
\usepackage{amsfonts}
\usepackage{amssymb}
\usepackage{graphicx}
\usepackage{tabularx}

\newcommand{\bra}[1]{\langle #1 |}
\newcommand{\ket}[1]{| #1 \rangle}

\newcommand{\mean}[1]{\langle #1 \rangle}

\newcommand{\Ham}{\mathcal{H}}
\newcommand{\norm}[1]{\| #1 \|}
\newcommand{\Id}{\mathds{1}}

\newcommand{\HILB}{\mathscr{H}}

\begin{document}
\newdimen\origiwspc%
\newdimen\origiwstr%
\preprint{}

\title{Full counting statistics approach to the quantum non-equilibrium Landauer bound}
\author{Giacomo Guarnieri,$^{1,2,3}$ Steve Campbell,$^{1,2,4}$ John Goold,$^5$ Simon Pigeon,$^{4,6}$ Bassano Vacchini,$^{1,2}$ and Mauro Paternostro$^4$}
\affiliation{
$^1$Dipartimento di Fisica, Universit{\`a} degli Studi di Milano, Via Celoria 16, 20133 Milan, Italy\\
$^2$Istituto Nazionale di Fisica Nucleare, Sezione di Milano, Via Celoria 16, 20133 Milan, Italy\\
$^3$Department of Optics, Palack\'{y} University, 17. listopadu 1192/12, 771 46 Olomouc, Czech Republic\\
\mbox{$^4$Centre for Theoretical Atomic, Molecular and Optical Physics, Queen's University Belfast, Belfast BT7 1NN, United Kingdom}\\
$^5$The Abdus Salam ICTP, Strada Costiera 11, 34151, Trieste, Italy\\
$^6$Laboratoire Kastler Brossel, UPMC-Sorbonne Universit\'es, CNRS, ENS-PSL Research University, Coll\`ege de France, 4 place Jussieu Case 74, F-75005 Paris, France
}
\begin{abstract}
We develop the full counting statistics of dissipated heat to explore the relation with Landauer's principle. Combining the two-time measurement protocol for the reconstruction of the statistics of heat with the minimal set of assumptions for Landauer's principle to hold, we derive a general one-parameter family of upper and lower bounds on the mean dissipated heat from a system to its environment. Furthermore, we establish a connection with the degree of non-unitality of the system's dynamics and show that, if a large deviation function exists as stationary limit of the above cumulant generating function, then our family of lower and upper bounds can be used to witness and understand first-order dynamical phase transitions. For the purpose of demonstration, we apply these bounds to an externally pumped three level system coupled to a finite sized thermal environment.
\end{abstract}
\date{\today}
\maketitle

\section{INTRODUCTION}
\label{sec:Introduction}

In his landmark 1961 paper, Rolf Landauer demonstrated that the heat dissipated in an irreversible computational process must always be at least equal to the corresponding information theoretic entropy change~\cite{Landauer1961}. A major implication of Landauer's principle, which is a fundamental statement on the energetic cost of information processing, is the resolution of Maxwell's daemon paradox~\cite{leffrex,Smoluchowski:1912,Szilard:29,Brillouin:1951,Penrose:1970,Bennett} that lurked in the background of statistical mechanics since its inception. 

The understanding of how a system dissipates heat following the manipulation of the information brought about by its relevant degrees of freedom is  important from both a fundamental and practical standpoint, in particular to gauge the energetics and thermodynamics of small classical and quantum systems. In fact, the miniaturization of technologies has led a significant interest in the thermodynamics of small systems that are out-of-equilibrium, both from the classical \cite{jrev,seifert} and quantum point of view~\cite{Esposito2009,mrev,JohnReview}. One the most exciting developments in this line of research is the recent availability of experimental platforms to explore energetic features of small information processing systems~\cite{Toyabe:2010,Orlov:2012,BerutNature,JunPRL,Koski1,Koski2,Gaudenzi}. In the quantum domain, Landauer's principle  has been studied extensively \cite{Lubkin, Plenio:2001, Maruyama,AndersPRE,MarcoNJP,RuPRL,OmarNJP}, and the first experiments addressing the energetic costs of information processing are just coming along~\cite{Peterson:2016,Vidrighin:2016,Ciampini,Mancino}. The ultimate limit of information-to-energy conversion set by Landauer's principle, including finite-size corrections due to the finite-size nature of the environment being addressed~\cite{EspositoNJP,ReebWolf}, was reached in an NMR setup implementing a two-qubit quantum gate~\cite{Peterson:2016} and following a proposal based on measuring the first moment of the statistics of heat exchanges~\cite{Goold:14}.

Recently, some of us studied a Landauer erasure process from the perspective of the full statistics of dissipated heat~\cite{MauroJohn}, showing that a novel lower bound can be derived which depends on the degree of non-unitality of the quantum operation induced on the environment (cf. related works \cite{Pillet2015,Bedingham:16}). In this paper we go beyond such an approach and apply the formalism of full counting statistics to dissipated heat in order to derive a new family of single-parameter lower (and upper) bounds on the average dissipated heat. Such a family of bounds can be made arbitrarily tight and does not depend on the details of the map, thus marking their inherent difference from the lower bound derived in Ref.~\cite{MauroJohn}, which is contained in our results as a particular case. We show how the bounds relate to a large deviation function, which is typically used for analyzing the long time statistical properties of a given system~\cite{Touchette}. In order to illustrate the behavior of the bounds thus derived, we make use of an engineered setting where a three-level system is coupled to a finite-dimensional thermal environment. While allowing for the demonstration of the tightness of the bound, such example allows us to shed light on the occurrence of interesting statistical phenomena such as dynamical phase transitions.

The remainder of the paper is organized as follows. In Sec.~\ref{sec:Formalism} we detail the formalism applied throughout this work. In Sec.~\ref{boundsderived} we derive the family of bounds and examine them through a large deviation approach. Sec.~\ref{physicalmodel} is dedicated to the behavior of the bounds with respect to a specific physical system. Finally, in Sec.~\ref{conclusionsect} we present our conclusions. Some technical details are outlined in Appendix~\ref{App:A}.

\section{FORMALISM}
\label{sec:Formalism}
\subsection{Erasure protocol}
\label{erasure}
Consider a system $\mathit{S}$ whose information content we want to erase by making it interact with an environment $\mathit{E}$. Following Refs.~\cite{Landauer1961,ReebWolf}, we consider the following minimal set of assumptions, which ensure the validity of Landauer's principle:
\begin{enumerate}
\item Both $\mathit{S}$ and $\mathit{E}$ are quantum systems, living in Hilbert spaces $\HILB_S$ and $\HILB_E$ respectively;
\item The initial state of the composite system is factorized, i.e. $\rho_{SE}(0) = \rho_S(0)\otimes\rho_E(0)$, such that no initial correlations are present;
\item The environment is prepared in the thermal state $\rho_E(0)=\rho_{\beta}=e^{-\beta\Ham_E}/Z_E$ with $\Ham_E$ the Hamiltonian of the environment, which we spectrally decompose as ${\cal H}_E= \sum_m E_m \ket{E_m}\bra{E_m} = \sum_m E_m \Pi_m$. Here, $\ket{E_m}$ is the $m^{\rm th}$ eigenstate of ${\cal H}_E$, associated with eigenvalue $E_m$. Finally, we have introduced the partition function $Z_E = \mathrm{Tr}_E\left[e^{-\beta\Ham_E}\right]$;
\item System and environment interact via the overall unitary transformation $U(t) = e^{-i\Ham t}$ with $\Ham = \Ham_S + \Ham_E + \Ham_{SE}$ the total Hamiltonian. 
\end{enumerate}
Within this framework, which is rather natural, the following equality has been proven~\cite{EspositoNJP,ReebWolf}
\begin{equation}
\beta \langle Q \rangle_t = \Delta S(t) + I(\rho_S(t) : \rho_E(t)) + D(\rho_E(t)||\rho_\beta),
\end{equation}
where $\langle Q \rangle_t \equiv \mathrm{Tr}\left[\Ham_E (\rho_E(t)-\rho_E(0))\right]$ is the mean dissipated heat, $\Delta S(t) \equiv S(\rho_S(0)) - S(\rho_S(t))$ is the change in the system's entropy (with $S(\rho) \equiv -\mathrm{Tr}\left[\rho\ln\rho\right]$ the von-Neumann entropy), $ D(\rho_E(t)||\rho_\beta) \equiv \mathrm{Tr}\left[\rho_E(t)\ln\rho_E(t)\right] - \mathrm{Tr}\left[\rho_E(t)\ln\rho_{\beta}\right] $ is the relative entropy between the state of the environment at time $t$ and its initial equilibrium state, and where $I(\rho_S(t) : \rho_E(t)) \equiv S(\rho_S(t)) + S(\rho_E(t)) - S(\rho_{SE}(t))$ denotes the mutual information between $S$ and $E$. As both the relative entropy and the mutual information are non-negative functions, one is immediately led to the following lower bound to the mean dissipated heat
\begin{equation} 
\label{LandauerBound}
\beta \langle Q \rangle_t \geq \Delta S(t),
\end{equation}
which is the well-known Landauer's principle.

\subsection{Full counting statistics approach to dissipated heat}
We rely on the full counting statistics~\cite{Esposito2009} of the dissipated heat, defined as the change in the environmental energy~\cite{ReebWolf,EspositoNJP}, in order to characterize its mean value. The probability distribution, $p_t(Q)$, to record a transferred amount of heat $Q$ can be formally defined in terms of the so-called two-time measurement protocol, introduced in Ref.~\cite{Lutz07} for the sake of determining the distribution of work resulting from a (unitary) perturbation of a system. In line with the framework defined above, assume $\mathit{S}$ to be initially uncorrelated with $\mathit{E}$, which is prepared in an equilibrium state. Therefore $\rho_{SE}(0) = \rho_S(0)\otimes\rho_{\beta}$ with $\left[ \mathcal{H}_E, \rho_{\beta}\right]=0$. A projection over one of the energy eigenstates of the environment at time $t=0$ is carried out, obtaining $E_n$ as an outcome. As a result, the total $S$-$E$ state is 
\begin{equation}
\rho'_{SE}(0)=\rho_S(0)\otimes\Pi_n.
\end{equation}
Immediately after the measurement, the interaction between $\mathit{S}$ and $\mathit{E}$ is switched on and the overall system undergoes a joint evolution up to a generic time $t$, when the interaction is switched off and a second projective measurement of the environmental energy is performed, this time obtaining an outcome $E_m$. After the second measurement, we have
\begin{equation}
\rho''_{SE}(t) = \frac{\Pi_{m} U(t)\rho'_{SE}(0)  U(t)^{\dagger} \Pi_{m}}{\mathrm{Tr}_{SE}\left[\Pi_{m} U(t)\rho'_{SE}(0)  U(t)^{\dagger}\right]} .
\end{equation}
It is worth stressing that the set of assumptions and steps used in the two-time measurement protocol is perfectly compatible with those required by the erasure process given in Sec.~\ref{erasure}. The joint probability to have obtained the two stated outcomes at times $0$ and $t$ respectively is given by the Born rule
\begin{equation}
\label{joint}
\begin{aligned}
P_t\left[E_m,E_n\right] = \mathrm{Tr}[ & \Pi_{m} U(t) \Pi_{n} \rho_S(0)\otimes\rho_{\beta} \Pi_{n} U^{\dagger}(t) \Pi_{m} ],
\end{aligned}
\end{equation}
from which the probability distribution $p_t(Q)$ follows as
\begin{equation}\label{probQ}
p_t(Q)=\sum_{E_n,E_m} \delta(Q - (E_m-E_n)) P_t\left[E_m,E_n\right].
\end{equation}
We introduce the cumulant generating function defined as the Laplace transform of the probability distribution
\begin{equation}\label{ThetaStart}
\Theta(\eta,\beta,t) \equiv \ln \langle  e^{-\eta Q} \rangle_t = \ln \int p_t(Q) e^{-\eta Q} dQ,
\end{equation}
which can be seen as the Wick rotated version of the usual definition given by the Fourier transform of $p_t(Q)$. The reason behind this choice will become clear in the following Section.
The cumulant of $n^{th}$-order is simply obtained by differentiation with respect to the real parameter $\eta$ as
\begin{equation}\label{Cumulants}
\langle Q^n \rangle_t = (-1)^n\frac{\partial^n}{\partial\eta^n}\Theta(\eta,\beta,t)|_{\eta=0}.
\end{equation}
Note that in the definition of the cumulant generating function we have explicitly written the dependence on the inverse temperature $\beta$ of the bath, which enters in the joint probability Eq.~\eqref{joint} through the initial environmental state $\rho_{\beta}$. The crucial point in using the full counting statistics approach is that the cumulant generating function introduced in Eq.~\eqref{ThetaStart} can be expressed as
\begin{equation}
\Theta(\eta,\beta,t) = \ln\big( \mathrm{Tr}_S\left[\rho_S(\eta,\beta,t)\right] \big),
\end{equation}
where
\begin{equation}
\rho_S(\eta,\beta,t) = \text{Tr}_E \left[ U_{\eta/2}(t)\rho_S(0)\otimes\rho_{\beta}U^{\dagger}_{\eta/2}(t) \right],
\end{equation}
with $U_{\eta/2}(t) \equiv e^{-(\eta/2)\Ham_E} U(t) e^{(\eta/2)\Ham_E}$. By invoking the same approximations and techniques used to derive a master equation for the density matrix of the system $\rho_S(t)$, one can obtain a new equation for $\rho_S(\eta,\beta,t)$~\cite{Esposito2009}. Solving this is a task with the same degree of complexity as accessing the dynamics of the reduced system. In what follows, we circumvent such a difficulty by deriving a family of bounds, both lower and upper, to $\left<Q\right>_t$ using the counting statistics arising from the two-time measurement protocol.

\section{Bounds on the mean dissipated heat} 
\label{boundsderived}
\subsection{Lower bounds}
In order to derive a lower bound for $\mean{Q}_t$, we consider the cumulant generating function of its probability distribution. Having it defined as in Eq. \eqref{ThetaStart}, we can apply H\"{o}lder's inequality to prove that $\Theta(\eta,\beta,t)$ is a convex function with respect to the counting parameter $\eta$~\cite{Rockafellar1970}. This condition can be equivalently expressed as~\cite{Touchette}
\begin{equation}\label{Convexity}
\Theta(\eta,\beta,t) \geq \eta \frac{\partial}{\partial\eta}\Theta(\eta,\beta,t)\big|_{\eta=0}.
\end{equation}
Combining Eq.~\eqref{Cumulants} and Eq.~\eqref{Convexity}, we obtain a one-parameter family of lower bounds for the mean dissipated heat $\mean{Q}_t$ reading
\begin{equation}
\label{bound}
\beta \mean{Q}_t \geq -\frac{\beta}{\eta}\Theta(\eta,\beta,t) \equiv \mathcal{B}^{\eta}_{\mathcal{Q}}(t)\quad (\eta > 0).
\end{equation}
 Eq.~\eqref{bound} is valid in the case of a generic erasure protocol and forms a central result of this work. 

We now look at the form taken by the bound for $\eta\!=\!\beta$ and show that the result of Ref.~\cite{MauroJohn} emerges. 
\GGrev{For this particular value of the counting field parameter, Eq~\eqref{ThetaStart} reduces to}
\begin{equation}\label{MJcase}
e^{\Theta(\beta,\beta,t)} = \mean{e^{-\beta Q}}_t,
\end{equation}
\GGrev{which can be seen to correspond to the same quantity considered in Ref.~\cite{MauroJohn}, i.e. the average exponentiated heat.} The bound in Ref.~\cite{MauroJohn} was shown to be related to the degree of non-unitality of the quantum operation acting on the environment, which governs the evolution of the environmental state. The unitality condition can be expressed as
\begin{equation}
\label{eq:nonU}
\sum_k A_k(t) A_k^{\dagger}(t) = \openone_E,
\end{equation} 
where $A_k(t) \equiv A_{ij}(t) = \sqrt{\lambda_j} \bra{i} U(t) \ket{j}$ denote the Kraus operators for the environment obtained from the usual evolution operator $U(t)$, $\lbrace\ket{j},\lambda_j\rbrace$ being the eigenstates and eigenvalues of the initial density matrix of the system, i.e. $\rho_S(0) = \sum_j \lambda_j \ket{j}\bra{j}$. To show this connection, we consider the expression of the cumulant generating function
\begin{equation}
\begin{aligned}
\Theta(\eta,\beta,t) &= \ln\mathrm{Tr}_{SE} \left[e^{-(\eta/2)\Ham_E} U(t) e^{(\eta/2)\Ham_E} \right. \times\\
&\left. \times \rho_{SE}(0) e^{(\eta/2)\Ham_E} U^{\dagger}(t) e^{-(\eta/2)\Ham_E}\right].
\end{aligned}
\end{equation}
Exploiting the cyclicity of the trace and the condition $\left[e^{(\eta/2)\Ham_E},\rho_{\beta}\right]=0$, it is straightforward to show that the latter can be equivalently expressed as
\begin{equation}\label{res}
\Theta(\eta,\beta,t)  = \ln\mathrm{Tr}_{E}\left[\rho_{\beta} \mathbf{A}^{\eta}(t)\right]
\end{equation}
with
\begin{equation}\label{Aeta}
\mathbf{A}^{\eta}(t) \equiv \mathrm{Tr}_S\left[ U_{\beta-\eta}(t) \, \left(\rho_S(0)\otimes \openone_E \right) \, U^\dag_{\beta-\eta}(t)\right],
\end{equation}
where $U_{\beta-\eta}(t) = e^{-(\eta-\beta)\Ham_E/2} U(t) e^{(\eta-\beta)\Ham_E/2}$ represents the evolution conditional on the two-time measurement of the environmental energy. Eq.~\eqref{res} remarks the role of the $\eta\!=\!\beta$ choice: for this value of the counting parameter we find that the operator defined in Eq.~\eqref{Aeta} reduces to
\begin{equation}
\label{NE}
\begin{aligned}
\mathbf{A}^{\beta}(t) &=\mathrm{Tr}_S\left[U(t)\rho_S(0)\otimes\Id_E U^{\dagger}(t)\right]  
\equiv \sum_k\!A_k(t) A_k^{\dagger}(t).
\end{aligned}
\end{equation} 
\GGrev{
Now that we have clarified the connection between the one-parameter family of lower bounds obtained in this work and the bound derived in Ref.~\cite{MauroJohn}, it is important to clarify the differences between the two techniques and the obtained results. Despite both approaches taking as a starting point the heat probability distribution $p_t(Q)$ given in Eq.~\eqref{probQ}, Ref.~\cite{MauroJohn} uses it to directly construct the average exponentiated heat of Eq.~\eqref{MJcase}, in the same spirit as Jarzynski for the case of the work probability distribution. This quantity does not allow one to obtain the moments of the distribution of dissipated heat by differentiation, and only the application of Jensen's inequality allows to access the lower bound on the mean dissipated heat given in Ref.~\cite{MauroJohn}. In our approach, instead one builds on the cumulant generating function, allowing to obtain both the different moments of the distribution, including in particular the mean values, as well as a family of upper and lower bounds to it. This last fact is of particular relevance in that it paves the way to assess the existence of dynamical phase transitions that will be explored in the proceeding sections.}

In light of Eqs.~\eqref{bound} and \eqref{eq:nonU}, one can see that, if the environmental map is unital,  the new family of lower bounds vanishes. However, in the erasure-protocol framework considered here the dynamical map $\Lambda_E : \rho_\beta \mapsto \rho_E(t)$ is, by construction, non-unital as the dissipative dynamics inevitably perturbs the initial Gibbs state of the environment in order to erase information stored in the system \cite{MauroJohn}. In order to relate these concepts more quantitatively, in Sec.~\ref{physicalmodel} we introduce the following figure of merit, which gives an estimate of the degree of non-unitality of a map
\begin{equation}
\label{nonunitalityE}
\mathcal{N}_E(t) = \norm{\mathbf{A}^{\beta}(t)-\openone_E},
\end{equation}
where $\norm{\cdot}$ denotes the Frobenius norm.

\subsection{Upper bounds and relation to the large deviation function}
\label{LDFsection}
Consider now, if it exists, the stationary limit
\begin{equation}
\label{LD}
\theta(\eta,\beta) \equiv \lim_{t\to +\infty} \Theta(\eta,\beta,t)/t,
\end{equation}
which is the so-called large-deviation function (LDF), a powerful theoretical tool widely employed in literature to access the statistical properties at long time-scales \cite{Touchette,GarrahanLesanovskyPRL,GarrahanPRA85,LebowitzSpohn,Paternostro1,LesanovskyPRL}. Moreover, the LDF can be associated to a specific evolution in the space of events, thus being equivalent to a free energy~\cite{GarrahanLesanovskyPRL}. The usual evolution for $\eta=0$ is called \textit{typical}, while for $\eta\neq 0$ is referred to as {\it rare}. In particular, discontinuities in $\theta(\eta,\beta)$ correspond to dynamical phase transitions. \GGrev{The bounds derived in Sec.~\ref{sec:Formalism} allows us to have a remarkably clear grasp on the connection between discontinuities in the LDF and dynamical phase transitions. To show this, consider again the convexity condition Eq.~\eqref{Convexity}. If we limit our attention to negative values of the counting parameter $\eta$, instead of Eq.~\eqref{bound} we obtain an upper bound for the dissipated heat in the form}
\begin{equation}
\label{upperbound}
\beta \mean{Q}_t \leq \frac{\beta}{|\eta|} \Theta(\eta,\beta,t) \equiv \tilde{\mathcal{B}}^{\eta}_{\mathcal{Q}}(t)\quad (\eta < 0).
\end{equation}
Clearly, this upper bound has similar properties as the lower bound found above, namely it approaches from above the curve of the dissipated heat for decreasing values of $|\eta|$. \GGrev{In light of this, it follows that if $\eta\!=\!0$ is a critical point for $\theta(\eta,\beta)$ (provided the long-time limit of Eq.~\eqref{LD} exists), then the two families }
\begin{equation}
b^{\eta}_{\mathcal{Q}} \equiv \lim_{t \to +\infty} \mathcal{B}^{\eta}_{\mathcal{Q}}(t)/t,\quad \tilde{b}^{\eta}_{\mathcal{Q}} \equiv \lim_{t \to +\infty}\tilde{\mathcal{B}}^{\eta}_{\mathcal{Q}}(t)/t
\end{equation}
approach two different curves, and thus provide a clear signature of a first-order dynamical phase coexistence in the typical evolution. If the critical point is instead located at some $\eta_c \neq 0$ it means the first order phase transition in the dissipated heat occurs for a rare evolution.

\section{Application to a Physical Model}
\label{physicalmodel}
\begin{figure}[t]
\begin{center}
\includegraphics[width=0.9\columnwidth]{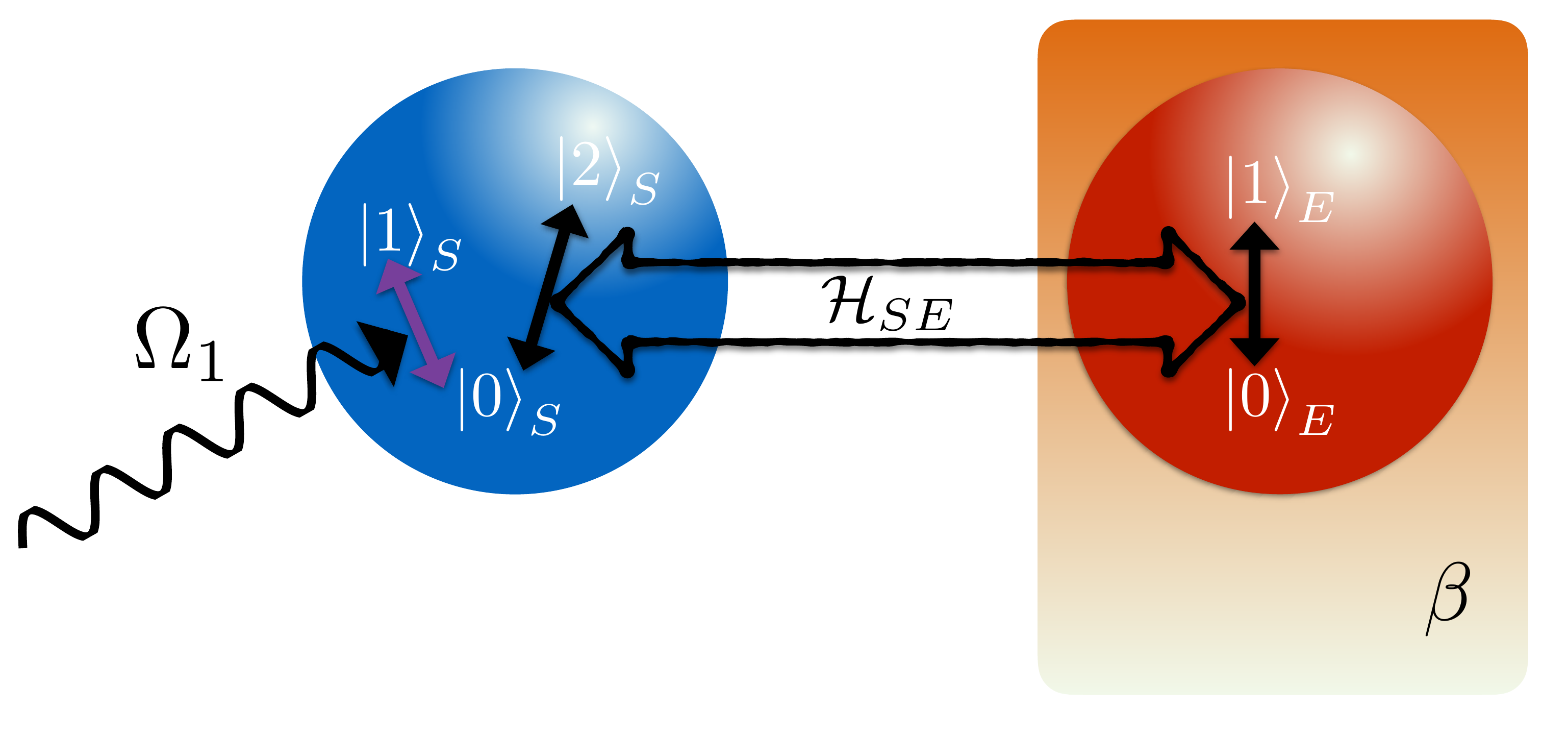}
\caption{Scheme of the physical system considered. A three-level V-system is coupled to a two level environment which is at thermal equilibrium and the $\ket{0}_S$- $\ket{1}_S$ transition is externally pumped.}
\label{fig1}
\end{center}
\end{figure}

Here we study the family of bounds in Eq.~\eqref{bound} in the context of a physical system consisting of a three level V-system encoded in the energy levels $\{ \ket{0}_S,~\ket{1}_S,~\ket{2}_S \}$ of a quantum system, such as in Fig.~\ref{fig1}. There, the $\ket{0}_S$-$\ket{1}_S$ transition is pumped with a frequency $\Omega_1$, while the transition between $\ket{0}_S$-$\ket{2}_S$ is dictated by an XX-type interaction with the environment, modelled as a two-level system (whose logical states are $\{ \ket{0}_E, \ket{1}_E \}$) and prepared in a thermal state. An external magnetic field along the $z$ direction affects both the environment and the $\ket{0}_S$-$\ket{2}_S$ transition. \GGrev{Such an effective model can arise considering a three level V-system in the context of adiabatic elimination~\cite{Sanz}}. As the interaction with the environment is excitation-preserving, the coupling behaves similarly to an amplitude damping channel affecting the $\ket{0}_S$-$\ket{2}_S$ transition, \GGrev{in fact it can be shown that the corresponding map applied to the $\ket{0}_S$-$\ket{2}_S$ transition is exactly a generalised amplitude damping channel~\cite{MauroJohn,BosePRL}.} This model thus shares many features with the one considered in Ref.~\cite{GarrahanLesanovskyPRL}, which was shown to exhibit a dynamical phase transition. 

Fig.~\ref{fig1} shows a schematic of the considered model. We will show the relation between the family of bounds in Eq.~\eqref{bound}, with particular emphasis applied to the special case of $\eta\!=\!\beta$, which matches the bounds derived in Ref.~\cite{MauroJohn}, and the actual dissipated heat. We will further show that the tightness of the lower bound can reveal characteristic features of the model and clearly explain the dynamics in light of the energy exchanged between system and environment.
Finally, the model considered provides a benchmark for the case of longer environmental chains. In fact, as highlighted in Ref.~\cite{MauroJohn}, the qualitative features of all the quantities of interest are already efficiently captured by the case of single-spin environment. 

\subsection{Coupled V-system}
The total Hamiltonian is given by $\mathcal{H}= \Ham_S + \Ham_E + \Ham_{SE} + \Ham_{SF}$ (where $F$ denotes the laser field), with 
\begin{align*}
&\GGrev{\Ham_S = - B S_z^{20},\quad \Ham_E = - B\sigma_z},\quad \Ham_{SF} = \Omega_1 S_+^{10} + \Omega_1^* S_-^{10},\\
&\Ham_{SE} = J\left( S_x^{20}\otimes\sigma_x  +S_y^{20}\otimes \sigma_y \right),
\end{align*}
where $\sigma_{x,y,z}$ denote the usual Pauli matrices for the environmental qubit, while $S_{x,y,z}^{j0}$ are the Gell-Mann matrices
\begin{equation}
\begin{aligned}
&S_x^{j0} \equiv \ket{0}\bra{j} + \ket{j}\bra{0}, \\
&S_y^{j0} \equiv i\left(\ket{0}\bra{j} - \ket{j}\bra{0}\right) ,\quad\quad (j=1,2)\\
&S_z^{j0} \equiv \ket{0}\bra{0} - \ket{j}\bra{j}.
\end{aligned}
\end{equation}
Finally $\Omega_1$ is the Rabi frequency of the $\ket{0}_S$-$\ket{1}_S$ transition and $S^{10}_{\pm} = \frac{1}{2}\left(S^{10}_x\pm iS^{10}_y\right)$. The evolution of the overall system can be analytically found and the solution, which is detailed in Appendix \ref{App:A}, puts into evidence the emergence of a typical frequency $\omega_1 = \sqrt{4J^2+\Omega_1^2}$, which plays a crucial role in the determination of many dynamical features, as shown below. In what follows, we will assume that the initial state is factorized as $ \rho(0) = \rho_S(0)\otimes\rho_{\beta}$, where $\rho_S(0) = \ket{2}_S\bra{2}$ and $\rho_{\beta}$ is a thermal state, in accordance with the assumptions made in the erasure protocol mentioned at the beginning of Sec.~\ref{erasure}.
\begin{figure}[t]
\begin{center}
\includegraphics[width=0.9\columnwidth]{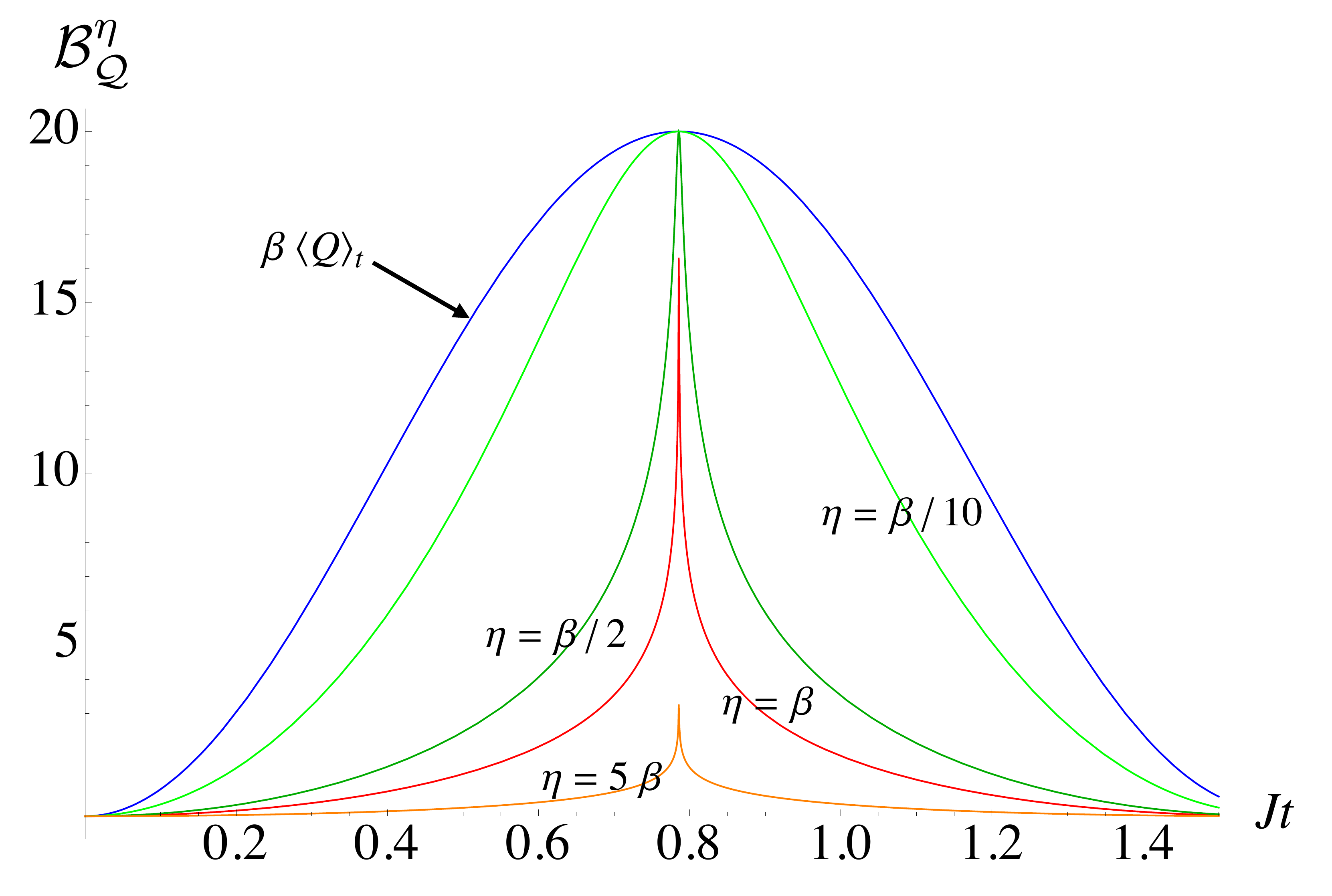}
\caption{Behaviour of the family of lower bounds, $\mathcal{B}^{\eta}_{\mathcal{Q}} $, for several values of $\eta$, fixing $B\!=\!1,~J\!=\!1,~\beta\!=\!10$ and $\Omega_1\!=\!0.1$. We also show the mean dissipated heat $\beta\mean{Q}_t$ (top-most blue curve) for reference. We remark the red curve at $\eta=\beta$ corresponds to the bound derived in Ref.~\cite{MauroJohn}.}
\label{fig2}
\end{center}
\end{figure}

\begin{figure}[t]
\begin{center}
{\bf (a)}\\
\includegraphics[width=0.9\columnwidth]{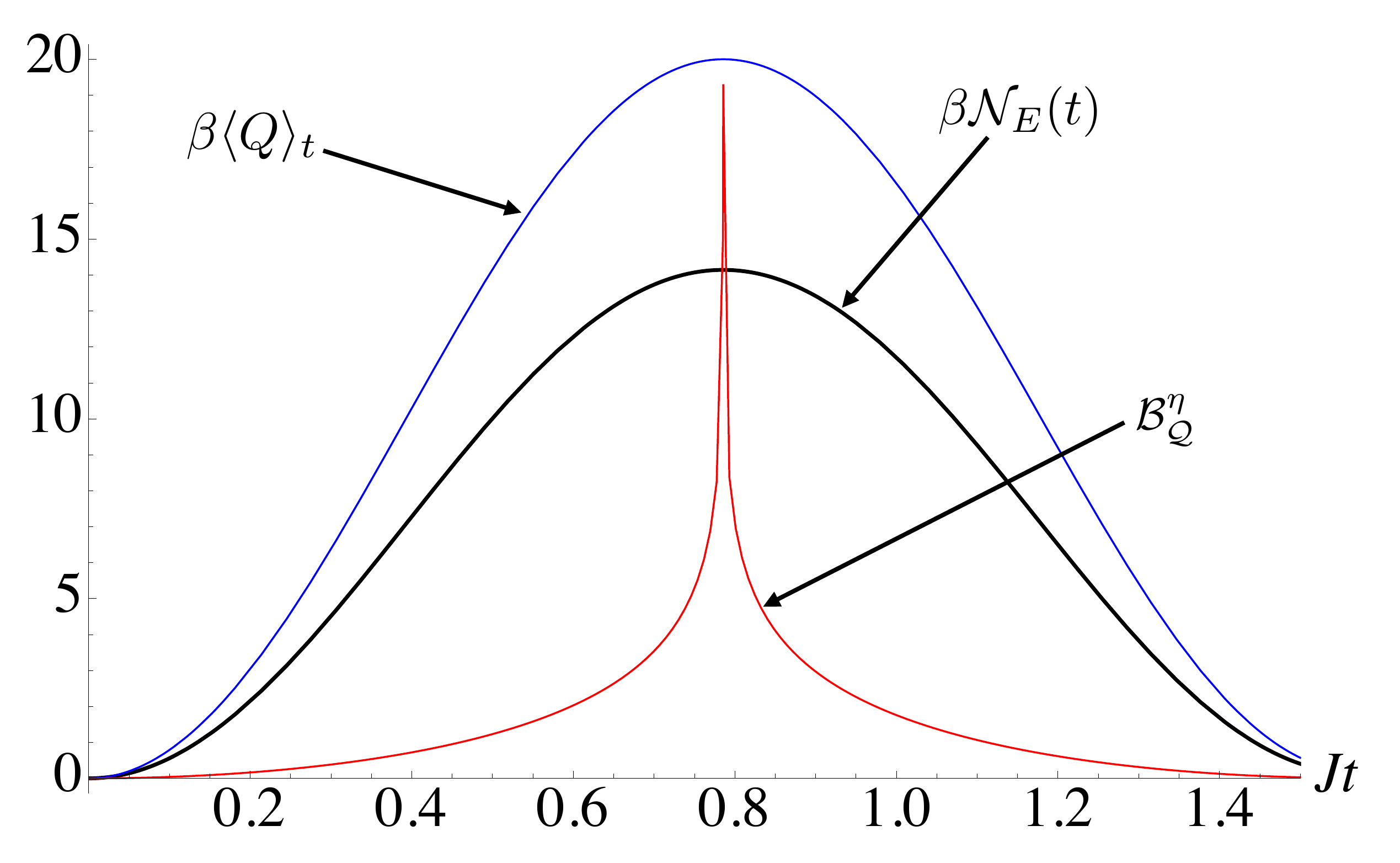}\\
{\bf (b)}\\
\includegraphics[width=0.9\columnwidth]{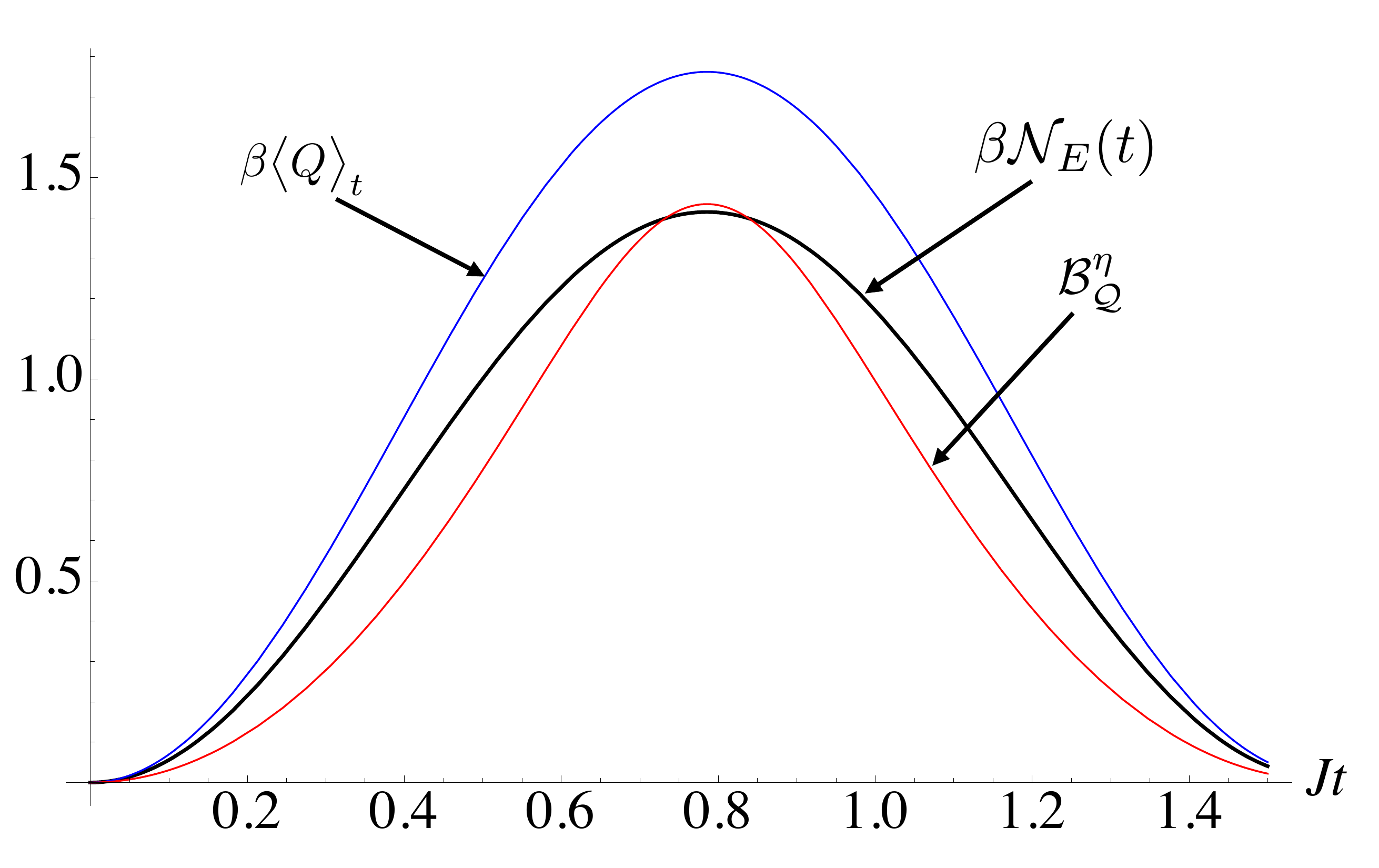}
\caption{Mean dissipated heat $\beta\mean{Q}_t$ (top-most blue curve), rescaled non-unitality $\beta\mathcal{N}_E(t)$ (middle, black curve) and the lower bound $\mathcal{B}^{\eta}_{\mathcal{Q}} $ for $\eta=\beta$ (bottom-most red curve). In both panels, we set $B=J=1$, $\Omega_1=0.1$ and take $\beta=10$ [$\beta=1$] in panel {\bf (a)} [{\bf (b)}].}
\label{fig3}
\end{center}
\end{figure}
\subsection{Behavior of the lower bounds}
Moving to the interaction picture with respect to the free Hamiltonian and employing the rotating-wave approximation, an analytic expression for the bounds $ \mathcal{B}^{\eta}_{\mathcal{Q}}(t) $ can be found. However, given their quite cumbersome nature, we refer the reader to Appendix~\ref{App:A}, focusing here only their behavior as a function of the dimensionless parameter $J t$. In Fig.~\ref{fig2} we (arbitrarily) fix $\beta\!=\!10$ and clearly see that, for decreasing values of the ratio $\eta/\beta$, the bound increasingly approaches the actual mean dissipated heat $\beta\mean{Q}_t$. We stress that the red line in Fig.~\ref{fig2}, corresponding to $\eta\!=\!\beta$, reproduces the lower bound obtained in Ref.~\cite{MauroJohn}. For larger values of $\eta/\beta$, the bound approaches zero. 

In Fig.~\ref{fig3} we show the behavior of the non-unitality measure $\mathcal{N}_E(t)$ defined in Eq.~\eqref{nonunitalityE} (rescaled with $\beta$), the dissipated heat, and the lower bound for $\eta\!=\!\beta$ in the cases of a cold and hot environmental state (corresponding to $\beta\!=\!10$ and $\beta\!=\!1$, respectively). Clearly, the zeros and maxima of the three curves are attained at the same times. A remarkable feature that occurs in Fig.~\ref{fig3} {\bf (a)} is the cusp appearing in $\mathcal{B}_\mathcal{Q}^\eta$, when the dissipated heat is maximized (the environmental qubit is effectively in the ground state as $\beta\!=\!10$). At the cusp, the bound is as close as possible to the actual dissipated heat. Contrarily, when $\beta\!=\!1$, such features are smoothed out and the dissipated heat is significantly reduced. Furthermore, the bound is now a smoothly varying function of the dimensionless time, closely tracking the functional form of $\beta \langle Q \rangle_t$ and $\beta \mathcal{N}_E(t)$. 

\begin{figure}[t]
\begin{center}
{\bf (a)}\\
\includegraphics[width=0.9\columnwidth]{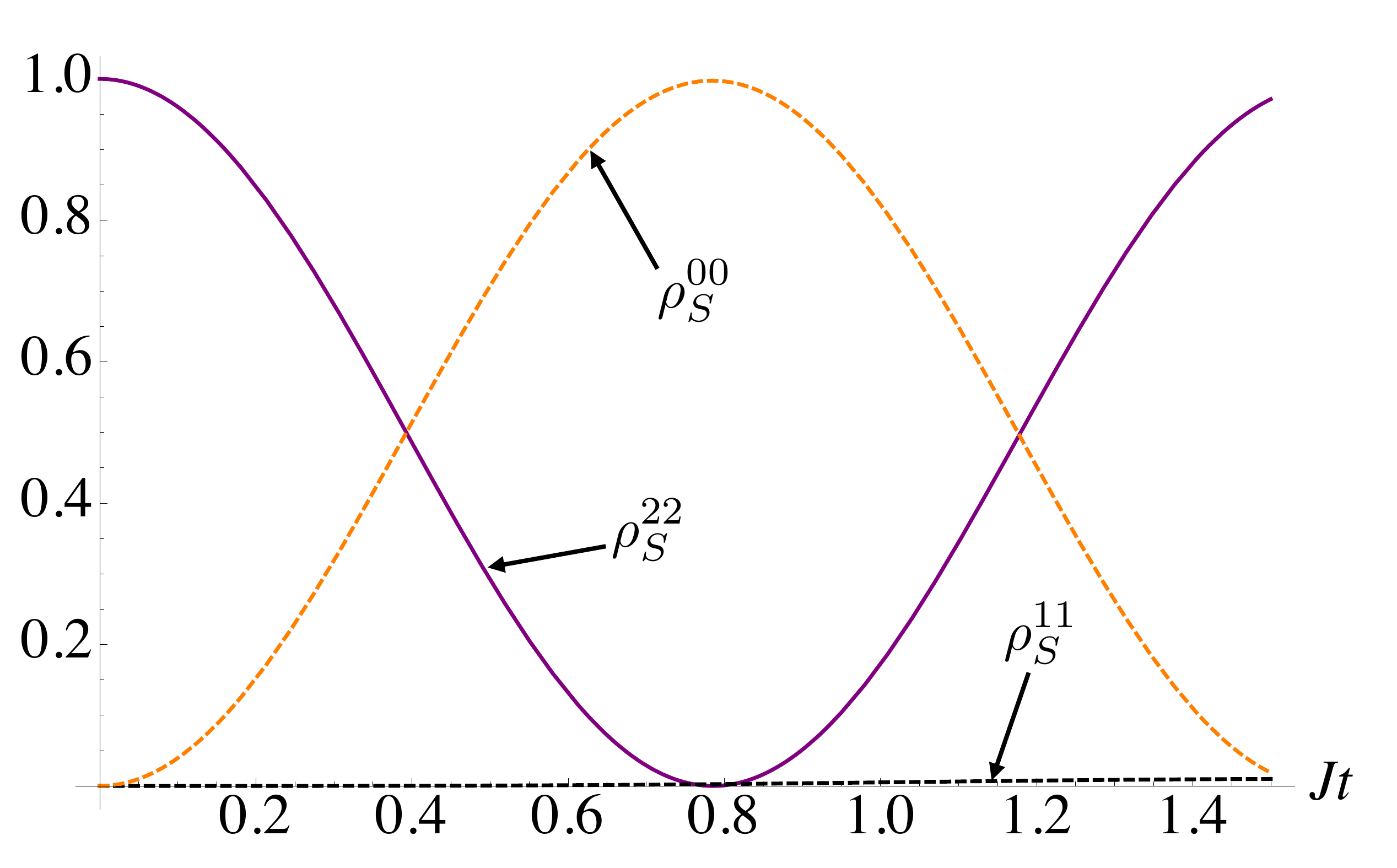}\\
{\bf (b)}\\
\includegraphics[width=0.9\columnwidth]{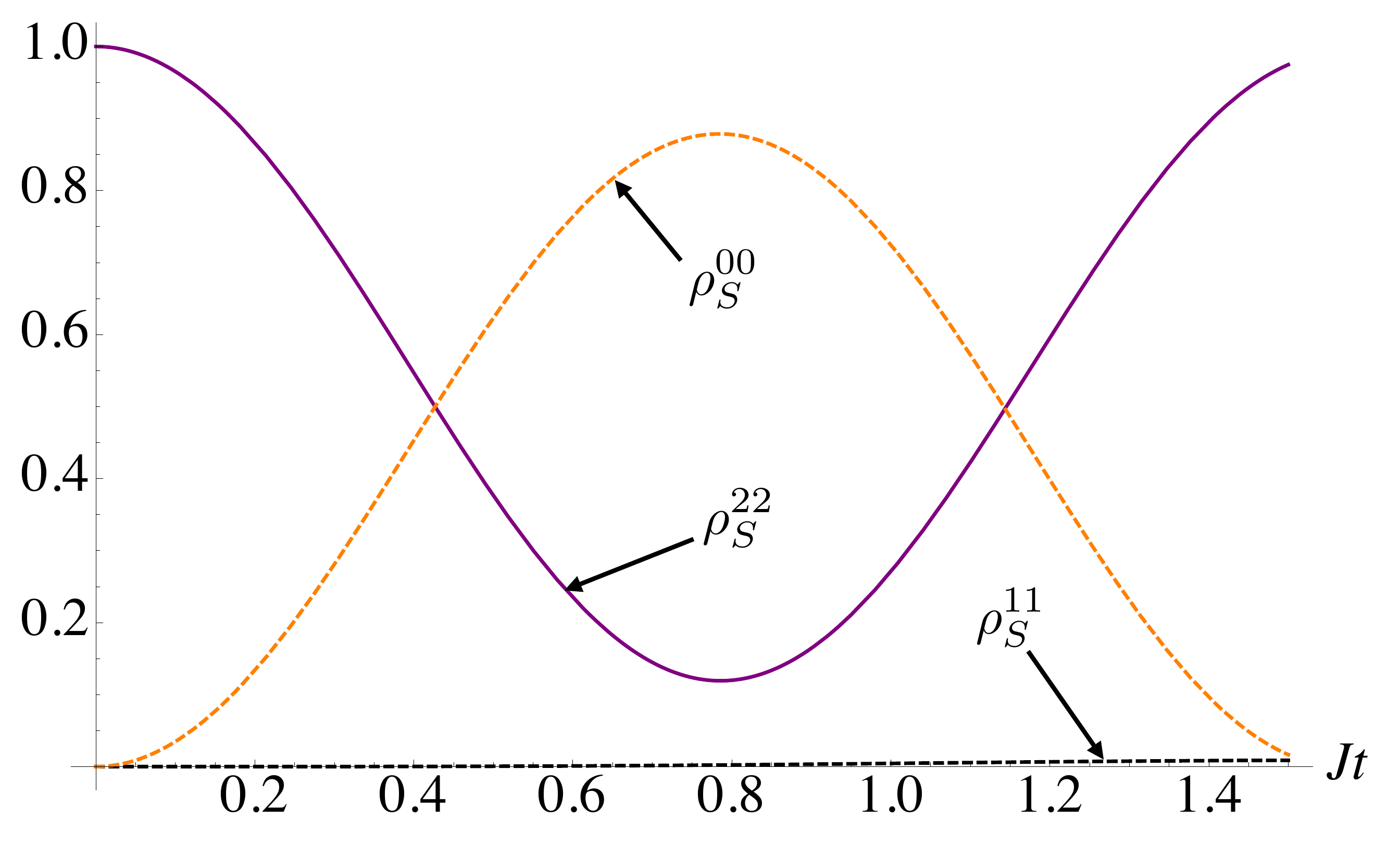}
\caption{Dynamics of the populations, $\rho_S^{00}$ (dashed, orange), $\rho_S^{11}$ (dashed, black), and $\rho_S^{22}$ (solid, purple). of the three-level system. {\bf (a)} For $B\!=\!1$, $J\!=\!1$, $\beta\!=\!10$ and $\Omega_1\!=\!0.1$. {\bf (b)} As for the previous panel except $\beta\!=\!1$.}
\label{fig4}
\end{center}
\end{figure}

This behavior can be explained by studying the populations, $\rho_S^{jj}~(j=0,1,2)$ of the V-system, shown in Fig.~\ref{fig4} for the same parameters used in Fig.~\ref{fig3}. Focusing on Fig.~\ref{fig4} {\bf (a)}, and recalling that we always assume our system is initialized in $\rho_S(0)=\ket{2}_S\bra{2}$, we see that as the system evolves the population of the $\ket{2}_S$ state is completely transferred to the $\ket{0}_S$ state. The point at which both $\beta\langle Q \rangle_t$ and $\mathcal{B}^\eta_\mathcal{Q}$ are maximized corresponds exactly to when $\rho_S^{22}=0$. At this point, all of the energy initially contained in the system is `dumped' into the environmental qubit, which was effectively in its ground state initially, and is thus able to absorb and store all of such energy. For $\beta=1$ [cf. Fig.~\ref{fig4} {\bf (b)}], the situation is markedly different due to the fact that the environment is comparatively warm, with a sizeable population initially in the excited state. In this case, the environment is unable to store {\it all} the energy initially in the system. Therefore state $\ket{2}_S$ cannot be depleted fully, and the dissipated heat is accordingly reduced.

\begin{figure}[t]
\begin{center}
\includegraphics[width=0.9\columnwidth]{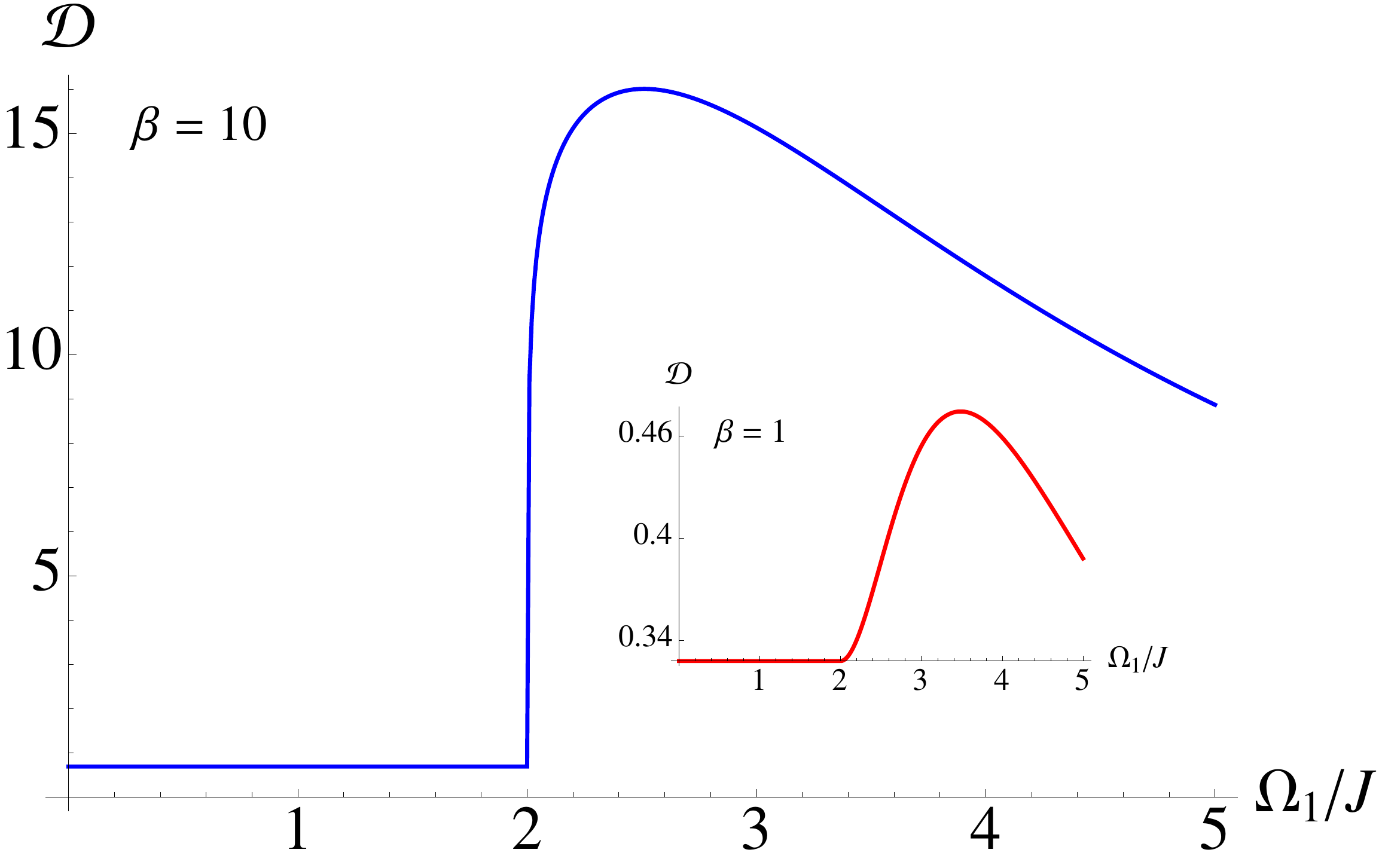}
\caption{Difference, $\mathcal{D}$, between the maximum of the mean dissipated heat and maximum of the lower bound $\mathcal{B}^{\eta}_{\mathcal{Q}}$ for $\eta=\beta$ as a function of the pump frequency $\Omega_1$. Here we take $J=1$, $B=1$, and $\beta=10$. {\it Inset:} As for main panel except setting $\beta=1$.}
\label{fig5}
\end{center}
\end{figure}

A closer examination of the cusp in Fig.~\ref{fig3} {\bf (a)} reveals a peculiar feature. By defining 
\begin{equation}
\label{difference}
\mathcal{D} = \max\left[\beta\langle Q \rangle_t \right] - \max\left[ \mathcal{B}^\eta_\mathcal{Q} \right],
\end{equation}
as the difference between the maximum dissipated heat and the maximum of the bound, we find that for the same parameters in Fig.~\ref{fig3} {\bf (a)}, $\mathcal{D} = \ln 2$. In Fig.~\ref{fig5} we provide a quantitative analysis of Eq.~\eqref{difference} to remark the existence of a `critical' pump strength. If the environment is initially cold (i.e. for $\beta=10$) and $\Omega_1/J\!\leq\!2$, we find that $\mathcal{D}=\ln 2$, exactly. This occurs because in this regime state $\ket{2}_S$ can always be fully emptied. If $\Omega\!>\!2J$, the pump starts dominating the dynamics. Due to the strong pumping of the $\ket{0}_S$-$\ket{1}_S$ transition, some of the population is trapped in the system and $\ket{2}_S$ is never completely empty. This induces a sudden increase in $\mathcal{D}$, due to the fact that, for $\Omega_1/J\!>\!2$, the bound is significantly reduced compared to the dissipated heat. Interestingly, the same qualitative behavior persists even when the environment is initially warm, i.e. for $\beta=1$. In the inset of Fig.~\ref{fig5} we see that for $\Omega_1/J\!\leq\!2$, $\mathcal{D}$ is again constant, and only changes when $\Omega_1/J\!>\!2$. 

\subsection{Dissipative regime: Upper bounds and LDF}
Finally, we examine the behavior in relation to the LDF. An exact master equation in the interaction picture governing the dynamics of the V-system can be written. In the case of an initially cold environment (i.e. for $\beta\to+\infty$), this takes the form 
\begin{align} \label{ME1}
\frac{d}{dt}&\rho(t) = -i\left[\tilde{\Ham}(t),\rho(t)\right] \notag\\
& + d_1(t) \left( G_-(t) \rho(t) G_-^{\dagger}(t) - \frac{1}{2} \lbrace G_-^{\dagger}(t) G_-(t), \rho(t) \rbrace\right) \notag\\
&+ d_2(t) \left( H_-(t) \rho(t) H_-^{\dagger}(t) -\frac{1}{2} \lbrace H_-^{\dagger}(t) H_-(t), \rho(t) \rbrace\right)
\end{align}
with $\tilde{\Ham}(t) = \Omega_1 \begin{pmatrix}
 0 & 0 & 0 \\
 0 & 0 & 1 \\
 0 & 1 & 0 \\
\end{pmatrix}$ and the parameters
\begin{equation}
\begin{aligned}
&d_{1,2}(t) = b(t)\mp\sqrt{b^2(t)+4a^2(t)},\quad \omega_1 \equiv \sqrt{4J^2+\Omega_1^2},\\
&a(t) = \frac{2 J^2\Omega_1 \left[1-\cos\left(\omega_1 t\right)\right]}{\omega_1^2-4J^2\left[1-\cos\left(\omega_1 t\right)\right]} ,\\
&b(t) = \frac{4 J^2 \omega_1\sin\left(\omega_1 t\right)}{\omega_1^2-4J^2\left[1-\cos\left(\omega_1 t\right)\right]}.
\end{aligned}
\end{equation}
The Lindblad operators  $G_-(t)$ and $H_-(t)$ are given by the following combinations of lowering operators 
\begin{equation}
\begin{aligned}
G_-(t) &= -v_-(t) \ket{1}_S\bra{2} + i\sqrt{1-v_-^2(t)}\ket{0}_S\bra{2}, \\
H_-(t) &= v_+(t) \ket{1}_S\bra{2} + i\sqrt{1-v_+^2(t)}\ket{0}_S\bra{2} 
\end{aligned}
\end{equation}
with $v_{\pm}(t) = \pm\frac{\sqrt{2} a(t)}{\sqrt{b(t) \left(b(t)\pm\sqrt{4 a^2(t)+b^2(t)}\right)+4 a^2(t)}}$. We stress that both $G_-(t)$ and $H_-(t)$ are normalized to $1$ and mutually orthogonal with respect to the Hilbert-Schmidt product, i.e. $\mathrm{Tr}_S\left[G_-^{\dagger} H_-(t)\right] = 0$. Note that, if we switch off the pump $\Omega_1$, the function $a(t)$ vanishes, while $b(t) \to 2J\tan\left(2Jt\right)$, and we thus get the following master equation
\begin{equation}
\label{standME}
\frac{d}{dt}\rho(t) = 2 J \tan\left(2Jt\right) \left( \sigma^{20}_- \rho(t) \sigma^{20}_+ -\frac{1}{2} \lbrace \sigma^{20}_+\sigma^{20}_-, \rho(t) \rbrace\right),
\end{equation}
which describes an amplitude-damping process involving the $\ket{0}_S$-$\ket{2}_S$ transition.

\begin{figure}[b]
\begin{center}
\includegraphics[width=0.9\columnwidth]{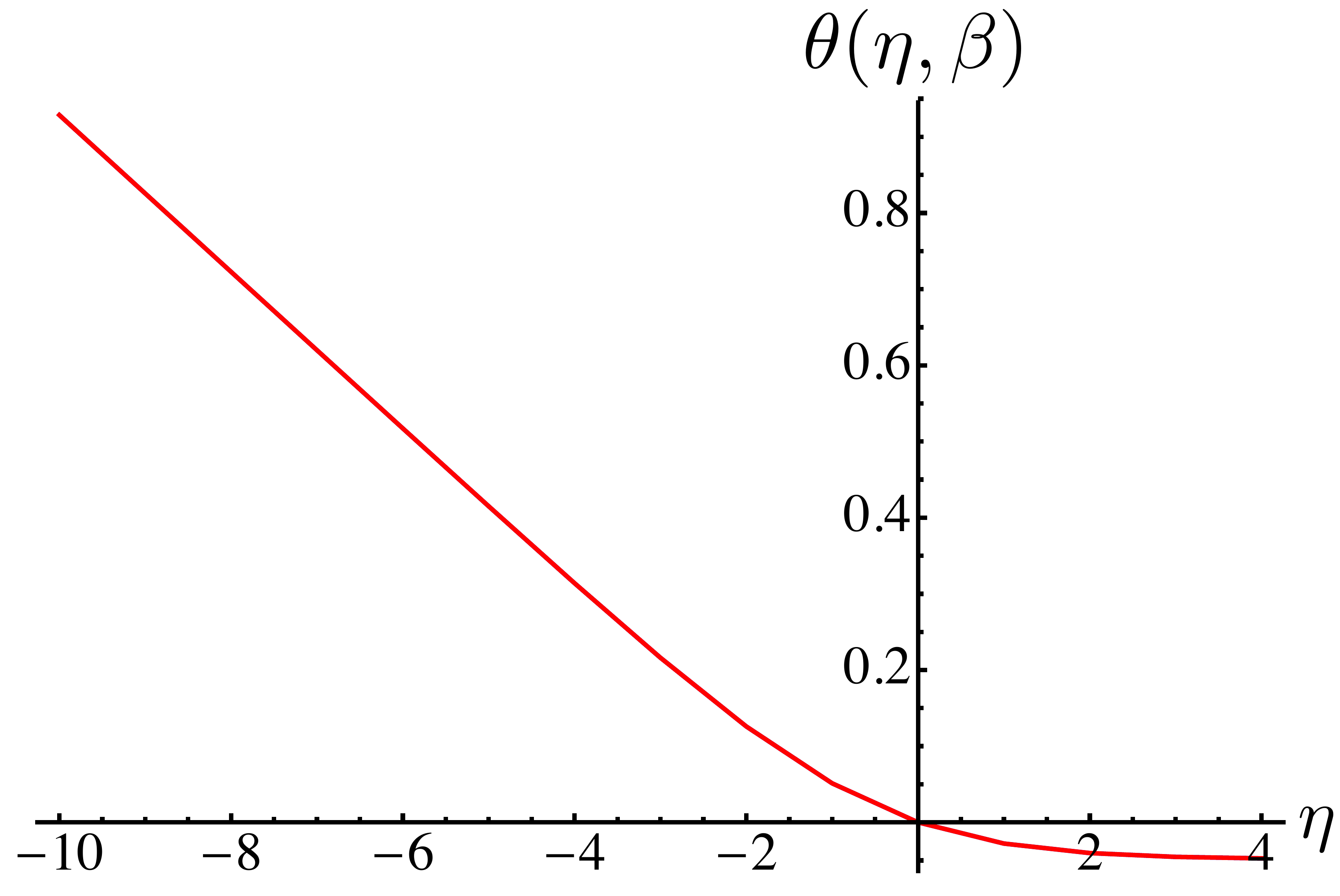}
\caption{LDF plotted against the counting parameter $\eta$ for $J=1$, $\gamma=4$, and $\Omega_1=0.01$. This behavior is valid for a zero-temperature environmental state. }
\label{fig7}
\end{center}
\end{figure}

Due to the finite size of the environment, the evolution of the system is periodic. More specifically, note that Eq.~\eqref{ME1} has the structure of a time-dependent Lindblad form and, although describing a completely-positive and trace-preserving channel, is not divisible. It thus describes a non-Markovian evolution even within a single period. Therefore, it is clear that the long-time limit in Eq.~\eqref{LD} does not exist. For this reason, we introduce an additional channel in the master equations of the form given by Eq.~\eqref{standME} with $2 J \tan\left(2Jt\right)\to\gamma$, which describes a decoherent interaction with an external bosonic field with a phenomenological damping constant $\gamma$. This proves sufficient to guarantee the existence of the large deviation function $\theta(\eta,\beta)$ which, when computed by numerical diagonalization, shows a crossover between two dynamical phases at $\eta=0$ [cf. Fig.~\ref{fig7}]. For $\eta<0$ the large deviation function becomes linear in $\eta$, while for $\eta>0$ quickly approaches a negative constant value determined by $\omega_1$. This result, which indicates a smoothed dynamical phase transition in the first moment of the dissipated heat, can be explained in the following way: for $\eta>0$ the three-level system evolves predominantly in the $\ket{0}_S$ - $\ket{1}_S$ subspace and correspondingly the dissipated heat, which is proportional to the derivative of the large deviation function, vanishes; for $\eta<0$, the dynamics involves instead the $\ket{0}_S$ - $\ket{2}_S$ transition which allows for energy to flow into the environmental spin, therefore leading to a dissipated heat. It is worth pointing out that the smoothness in the crossover between the two different dynamical phases takes into account the fact that the $\ket{0}_S$ level is, in the considered V-structure of the three-level system, shared by the two transitions, this therefore resulting in a non-vanishing probability to smoothly move from one phase to the other \GGrev{due to the external pump $\Omega_1$.
In the limiting case where the laser pump is switched off, i.e. $\Omega_1 \to 0$, the system undergoes a proper first-order dynamical phase transition, as reflected in a discontinuity in the first derivative of the LDF at the origin $\eta=0$.}

\section{Conclusions}
\label{conclusionsect}
We have presented a method to derive a one-parameter family of Landauer-like bounds for the mean dissipated heat based on the two-time measurement protocol. These bounds depend on the counting parameter $\eta$, and we have shown that they can be made arbitrarily tight. Remarkably, for $\eta\!=\!\beta$, the derived bound is exactly equal to the non-equilibrium lower bound derived by studying the dynamical map and employing a heat fluctuation relation~\cite{MauroJohn}. Applying these bounds to an interesting, yet simple, physical system, namely a pumped three level V-system coupled to a finite sized thermal environment, we showed how their tightness could highlight certain features of the system, in our case the emergence of a characteristic pumping frequency. We also introduced a clear qualitative relation between the mean dissipated heat, its lower bounds, and the degree of non-unitality of the governing dynamical map. Finally, we showed the formalism developed here could also be applied to the large deviation function analysis useful in studying dynamical phase transitions due to the fact it allows to obtain both upper and lower bounds on the mean dissipated heat. In light of the generality of the methodology employed, we expect our results to be applicable to other thermodynamically relevant quantities as well, such as work or entropy production. 

\acknowledgements
We acknowledge support from the EU Collaborative projects QuProCS (grant agreement 641277) and TherMiQ (grant agreement 618074), the UniMi H2020 Transition Grant, the Julian Schwinger Foundation (grant number JSF-14-7-0000), and the ANR ACHN C-Flight. G.G. acknowledges the support of the Czech Science Foundation (GACR grant no. GB14-36681G). MP is supported by the DfE-SFI Investigator Programme (grant 15/IA/2864) and the Royal Society Newton Mobility Grant NI160057. This work was partially supported by the COST Action MP1209.

\onecolumngrid
\appendix
\section{Details on the coupled V-system}
\label{App:A}
Here we provide a detailed discussion on the physical model considered in the main body of the work. First of all, it is important to notice that the cumulant generating function $\Theta(\eta,\beta,t)$ is left invariant by the passage to the interaction picture. This can be easily seen by taking into account the definition of $\Theta(\eta,\beta,t)$ and exploiting the cyclicity of the trace and the relation $\left[\Ham_0 , \Ham_E\right]=0$:
\begin{align}
\label{equivalencepictures}
\Theta(\eta,\beta,t) &= \ln\mathrm{Tr}_{SE} \left[e^{-(\eta/2)\Ham_E} U_0(t) U_I(t) e^{(\eta/2)\Ham_E} \rho_{SE}(0) e^{(\eta/2)\Ham_E} U_I^{\dagger}(t) U_0^{\dagger}(t) e^{-(\eta/2)\Ham_E}\right] \notag\\
&= \ln\mathrm{Tr}_{SE} \left[e^{-(\eta/2)\Ham_E} U_I(t) e^{(\eta/2)\Ham_E} \rho_{SE}(0) e^{(\eta/2)\Ham_E} U_I^{\dagger}(t) e^{-(\eta/2)\Ham_E}\right].
\end{align}
The above identity guarantees that we are free to move to the interaction picture with respect to the free Hamiltonian $\Ham_0 = \Ham_S + \Ham_E$ in order to access the full statistics of the dissipated heat.
Using the rotating wave approximation, the sum of the Hamiltonian contributions $\Ham_{SE} + \Ham_{SF}$ in the interaction picture reads
\begin{equation}\label{intHam}
\Ham_{SE}(t) + \Ham_{SF}(t) = J\left( S_x^{20}\otimes\sigma_x  +S_y^{20}\otimes \sigma_y \right) + \Omega_1 S_+^{10} + \Omega_1^* S_-^{10},
\end{equation}
The phase of the external field will be chosen in order for $\Omega_1$ to be real, so that
\begin{equation}\label{eq:HamInt}
\Ham_{SE}(t) + \Ham_{SF}(t) = J\left( S_x^{20}\otimes\sigma_x  +S_y^{20}\otimes \sigma_y \right) + \Omega_1 S_x^{10}\otimes\Id_2.
\end{equation}
This expression for the Hamiltonian in the interaction picture is then employed to obtain the cumulant generating function and subsequently the family of lower bounds $\mathcal{B}^{\eta}_{\mathcal{Q}}(t)$ using Eq.\eqref{equivalencepictures}. 

Moreover, if we assume the initial state to be of factorized form $ \rho(0) = \rho_S(0)\otimes\rho_{\beta}$, where $\rho_S(0) = \ket{\Psi_0}\bra{\Psi_0}$ with $\ket{\Psi_0} = \cos(\phi)\ket{0}_S + \sin(\phi)\sin(\alpha)\ket{1}_S + \sin(\phi)\cos(\alpha)\ket{2}_S$ and where$\rho_{\beta} = p\ket{0}_E\bra{0}+(1-p)\ket{1}_E\bra{1}$ with $p=\frac{1}{2}\left(1+\tanh (\beta B)\right)$ (also in accordance with the assumptions made in the erasure protocol mentioned at the beginning of Sec.\ref{sec:Formalism}), an exact master equation in the interaction picture can be found. For an initially cold environment (case $\beta\to+\infty$) the latter has the form given in Eq. \eqref{ME1}. Moreover, the cumulant generating function can be found analytically, though its expression for a generic choice of initial state of the system is quite cumbersome. For this reason, we report it below for the specific choice of $\theta=0, \phi=\pi/2$ which corresponds to $\rho_S(0) = \ket{2}_S\bra{2}$ considered in the main text:
\begin{align}
\Theta(\eta,\beta,t)\!=\!\log\!\!\left(\left[1\!+\tanh (\beta  B)\right]\!\frac{16 J^2 \Omega_1^2 e^{-2 B \eta } \sin ^4\left(\frac{\omega_1}{2} t\right)\!+\!4 J^2 e^{-2 B\eta } \sin ^2\left(\omega_1t\right)\!+\!\left(4 J^2 \cos \left(\omega_1t\right)+\Omega _1^2\right){}^2}{2\omega_1^4}\!+\!\frac{1\!-\tanh (\beta  B)}{2}\right),
\end{align}
where $\omega_1 = \sqrt{\Omega_1^2+4J^2}$. The family of lower bounds $\mathcal{B}^{\eta}_{\mathcal{Q}}(t)$ is then straightforwardly obtained. The mean dissipated heat $\beta\mean{Q}_t$ (blue line in Fig. \ref{fig1}) can be found analytically and reads
\begin{equation}
\label{meanQ1}
\mean{Q}_t = \left[1+\tanh (\beta  B)\right]\frac{16 B J^2 \sin ^2\left(\frac{\omega_1}{2} t\right) \left[\omega_1^2 - 4 J^2
   \sin^2 \left(\frac{\omega_1}{2} t\right)\right]}{\omega_1^4}.
\end{equation}
Note that this quantity is always positive for every value of the parameters $J, \Omega_1, B$ and at every time $t$.

Finally, the quantifier of the non-unitality degree of the environmental channel $\mathcal{N}_E(t) = \norm{\sum_{k} A_k(t)A_k^{\dagger}(t) - \Id_E}$ can be analytically accessed for this model. By direct exponentiation of the Hamiltonian \eqref{eq:HamInt}, the overall unitary evolution operator $U(t)$ governing the evolution of the composite system can in fact be found and reads 
\begin{equation}\label{U}
U(t) = \begin{pmatrix}
 1 & 0 & 0 & 0 & 0 & 0 \\
 0 & \frac{4 \cos \left(t \omega _1\right) J^2+\Omega _1^2}{\omega _1^2} & \frac{2
   J \left(\cos \left(t \omega _1\right)-1\right) \Omega _1}{\omega _1^2} & 0 &
   -\frac{2 i J \sin \left(t \omega _1\right)}{\omega _1} & 0 \\
 0 & \frac{2 J \left(\cos \left(t \omega _1\right)-1\right) \Omega _1}{\omega
   _1^2} & \frac{4 J^2+\cos \left(t \omega _1\right) \Omega _1^2}{\omega _1^2} & 0
   & -\frac{i \sin \left(t \omega _1\right) \Omega _1}{\omega _1} & 0 \\
 0 & 0 & 0 & \cos \left(t \Omega _1\right) & 0 & -i \sin \left(t \Omega _1\right)
   \\
 0 & -\frac{2 i J \sin \left(t \omega _1\right)}{\omega _1} & -\frac{i \sin
   \left(t \omega _1\right) \Omega _1}{\omega _1} & 0 & \cos \left(t \omega
   _1\right) & 0 \\
 0 & 0 & 0 & -i \sin \left(t \Omega _1\right) & 0 & \cos \left(t \Omega _1\right)
   \\
\end{pmatrix},
\end{equation}
from which the Kraus operators for the environmental channel $A_k$ can be found simply by taking the partial trace over the system. Note that the above expression Eq. \eqref{U} is given by assuming the following lexicographic order to expand the vectors $\ket{\Psi}\in\HILB_S\otimes\HILB_E = (\ket{21},\ket{20},\ket{11},\ket{10},\ket{01},\ket{00})^T$, where the first digit refers to the  the V-system while the second to the environmental qubit. 

The Frobenius norm of the difference between $\mathbf{A}^{\beta}(t)$ and the identity $\Id_E$ can be expressed in a closed form which, in the case $\rho_S(0) = \ket{2}_S\bra{2}$, reduces to
\begin{equation}
\mathcal{N}_E(t) =  \frac{16 \sqrt{2} J^2 \sin ^2\left(\frac{\omega_1}{2} t \right) \left[\omega_1^2 - 4 J^2 \sin^2 \left(\frac{\omega_1}{2}t\right)\right]}{\omega_1^4} = \frac{\sqrt{2}}{1+\tanh (\beta B)} \mean{Q}_t,
\end{equation}
We point out that this last result, which clearly shows the link between the mean dissipated heat and the degree of non-unitality, does not hold in general for a generic initial state of the system, but only for the choice $\rho_S(0) = \ket{2}_S\bra{2}$. In general however, this quantity is always positive and vanishes whenever the coupling $J$ goes to zero, or whenever the argument of the sine term goes to zero.

\twocolumngrid

\end{document}